
\input phyzzm


\catcode`@=11 

%
\paperstyle   
%
%
\def\MEMO{\letterstyle\FRONTPAGE \letterfrontheadline={\hfil}
    \line{\quad\fourteenrm MEMORANDUM\hfil\twelverm\the\date\quad}
    \medskip \memod@f}

\def\memit@m#1{\smallskip \hangafter=0 \hangindent=1in
      \Textindent{\caps #1}}
\def\memod@f{\xdef\to{\memit@m{To:}}\xdef\from{\memit@m{From:}}%
     \xdef\topic{\memit@m{Topic:}}\xdef\subject{\memit@m{Subject:}}%
     \xdef\rule{\bigskip\hrule height 1pt\bigskip}}
\memod@f
\newskip\lettertopfil
\lettertopfil = 0pt plus 1.5in minus 0pt
\newskip\letterbottomfil
\letterbottomfil = 0pt plus 2.3in minus 0pt
\newskip\spskip \setbox0\hbox{\ } \spskip=-1\wd0
\def\addressee#1{\medskip\rightline{\the\date\hskip 30pt} \bigskip
   \vskip\lettertopfil
   \ialign to\hsize{\strut ##\hfil\tabskip 0pt plus \hsize \cr #1\crcr}
   \medskip\noindent\hskip\spskip}
\newskip\signatureskip       \signatureskip=40pt
\def\signed#1{\par \penalty 9000 \bigskip \dt@pfalse
  \everycr={\noalign{\ifdt@p\vskip\signatureskip\global\dt@pfalse\fi}}
  \setbox0=\vbox{\singlespace \halign{\tabskip 0pt \strut ##\hfil\cr
   \noalign{\global\dt@ptrue}#1\crcr}}
  \line{\hskip 0.5\hsize minus 0.5\hsize \box0\hfil} \medskip }

\def\endletter{\ifnum\pagenumber=1 \vskip\letterbottomfil\supereject
\else \vfil\supereject \fi}
\newbox\letterb@x
\def\lettertext{\par\unvcopy\letterb@x\par}
\def\multiletter{\setbox\letterb@x=\vbox\bgroup
      \everypar{\vrule height 1\baselineskip depth 0pt width 0pt }
      \singlespace \topskip=\baselineskip }
\def\letterend{\par\egroup}
%
%
%
\newskip\frontpageskip
\newtoks\pubtype
\newtoks\Pubnum
\newtoks\pubnum
\newif\ifp@bblock  \p@bblocktrue
\def\PH@SR@V{\doubl@true \baselineskip=24.1pt plus 0.2pt minus 0.1pt
             \parskip= 3pt plus 2pt minus 1pt }
\def\PHYSREV{\paperstyle\PhysRevtrue\PH@SR@V}
\def\titlepage{\FRONTPAGE\paperstyle\ifPhysRev\PH@SR@V\fi
   \ifp@bblock\p@bblock\fi}
\def\nopubblock{\p@bblockfalse}
\def\endpage{\vfil\break}
\frontpageskip=1\medskipamount plus .5fil
\pubtype={ }
\newtoks\publevel
\publevel={Report}   
\Pubnum={}
%
\def\p@bblock{\begingroup \tabskip=\hsize minus \hsize
   \baselineskip=1.5\ht\strutbox \topspace-2\baselineskip
   \halign to\hsize{\strut ##\hfil\tabskip=0pt\crcr
   \the\Pubnum\cr \the\date\cr }\endgroup}
\def\title#1{\vskip\frontpageskip \titlestyle{#1} \vskip\headskip }
\def\author#1{\vskip\frontpageskip\titlestyle{\twelvecp #1}\nobreak}

\def\address#1{\par\kern 5pt\titlestyle{\twelvepoint\it #1}}
\def\andaddress{\par\kern 5pt \centerline{\sl and} \address}


%
\def\abstract{\vskip\frontpageskip\centerline{\fourteenrm ABSTRACT}
              \vskip\headskip }

%
%
%

\def\\{\relax\ifmmode\backslash\else$\backslash$\fi}
\def\globaleqnumbers{\relax\if\equanumber<0\else\global\equanumber=-1\fi}

\def\journal#1&#2(#3){\unskip, \sl #1~\bf #2 \rm (19#3) }

\def\topspace{\hrule height 0pt depth 0pt \vskip}
\def\prop{\mathrel{{\mathchoice{\pr@p\scriptstyle}{\pr@p\scriptstyle}{
                \pr@p\scriptscriptstyle}{\pr@p\scriptscriptstyle} }}}
\def\pr@p#1{\setbox0=\hbox{$\cal #1 \char'103$}
   \hbox{$\cal #1 \char'117$\kern-.4\wd0\box0}}
\def\lsim{\mathrel{\mathpalette\@versim<}}
\def\gsim{\mathrel{\mathpalette\@versim>}}
\def\@versim#1#2{\lower0.2ex\vbox{\baselineskip\z@skip\lineskip\z@skip
  \lineskiplimit\z@\ialign{$\m@th#1\hfil##\hfil$\crcr#2\crcr\sim\crcr}}}
%
%
%
\let\sec@nt=\sec
\def\sec{\relax\ifmmode\let\n@xt=\sec@nt\else\let\n@xt\section\fi\n@xt}
\def\obsolete#1{\message{Macro \string #1 is obsolete.}}
\def\firstsec#1{\obsolete\firstsec \section{#1}}
\def\firstsubsec#1{\obsolete\firstsubsec \subsection{#1}}
\def\thispage#1{\obsolete\thispage \global\pagenumber=#1\frontpagefalse}
\def\thischapter#1{\obsolete\thischapter \global\chapternumber=#1}
\def\nextequation#1{\obsolete\nextequation \global\equanumber=#1
   \ifnum\the\equanumber>0 \global\advance\equanumber by 1 \fi}
\def\BOXITEM{\afterassigment\B@XITEM\setbox0=}
\def\B@XITEM{\par\hangindent\wd0 \noindent\box0 }
%

%
\catcode`@=12 

%
%
\catcode`@=11
%
%

%
\font\fourteentt=cmtt10 scaled\magstep2  
\def\seventeenpoint{\relax
    \textfont0=\seventeenrm         \scriptfont0=\twelverm
    \scriptscriptfont0=\tenrm
     \def\rm{\fam0 \seventeenrm \f@ntkey=0 }\relax
    \textfont1=\seventeeni          \scriptfont1=\twelvei
    \scriptscriptfont1=\teni
     \def\oldstyle{\fam1 \seventeeni\f@ntkey=1 }\relax
    \textfont2=\seventeensy         \scriptfont2=\twelvesy
    \scriptscriptfont2=\tensy
    \textfont3=\seventeenex     \scriptfont3=\seventeenex
    \scriptscriptfont3=\seventeenex
    \def\it{\fam\itfam \seventeenit\f@ntkey=4 }
         \textfont\itfam=\seventeenit
    \def\sl{\fam\slfam \seventeensl\f@ntkey=5 }
         \textfont\slfam=\seventeensl
    \scriptfont\slfam=\twelvesl
    \def\bf{\fam\bffam \seventeenbf\f@ntkey=6 }
         \textfont\bffam=\seventeenbf
    \scriptfont\bffam=\twelvebf  \scriptscriptfont\bffam=\tenbf
    \def\tt{\fam\ttfam \fourteentt \f@ntkey=7 }
         \textfont\ttfam=\fourteentt
    \h@big=11.9\p@{} \h@Big=16.1\p@{} \h@bigg=20.3\p@{} \h@Bigg=24.5\p@{}
    \setbox\strutbox=\hbox{\vrule height 12pt depth 5pt width\z@}
    \samef@nt}
%

%

%
%
\mathchardef\bfalpha  ="090B
\mathchardef\bfbeta   ="090C
\mathchardef\bfgamma  ="090D
\mathchardef\bfdelta  ="090E
\mathchardef\bfepsilon="090F
\mathchardef\bfzeta   ="0910
\mathchardef\bfeta    ="0911
\mathchardef\bftheta  ="0912
\mathchardef\bfiota   ="0913
\mathchardef\bfkappa  ="0914
\mathchardef\bflambda ="0915
\mathchardef\bfmu     ="0916
\mathchardef\bfnu     ="0917
\mathchardef\bfpi     ="0918
\mathchardef\bfrho    ="091A
\mathchardef\bfsigma  ="091B
\mathchardef\bftau    ="091C
\mathchardef\bfupsilon="091D
\mathchardef\bfphi    ="091E
\mathchardef\bfchi    ="091F
\mathchardef\bfpsi    ="0920
\mathchardef\bfomega  ="0921
%
%
\newtoks\heth
\newtoks\Heth
\Pubnum={KUNS \the\pubnum}
\Heth={HE(TH)\the\heth}
\date={\monthname,\ \number\year}
\pubnum={000}
\heth={00/00}
\def\titlepage{\FRONTPAGE\ifPhysRev\PH@SR@V\fi
   \ifp@bblock\p@bblock\fi}
\def\p@bblock{\begingroup \tabskip=\hsize minus \hsize
   \baselineskip=1.5\ht\strutbox \topspace-2\baselineskip
   \halign to\hsize{\strut ##\hfil\tabskip=0pt\crcr
   \the\Pubnum\cr \the\Heth\cr \the\date\cr }\endgroup}
\def\titlestyleb#1{\par\begingroup \interlinepenalty=9999
     \leftskip=0.02\hsize plus 0.23\hsize minus 0.02\hsize
     \rightskip=\leftskip \parfillskip=0pt
     \hyphenpenalty=9000 \exhyphenpenalty=9000
     \tolerance=9999 \pretolerance=9000
     \spaceskip=0.333em \xspaceskip=0.5em
     \iftwelv@\fourteenpoint\else\twelvepoint\fi
   \noindent {\bf #1}\par\endgroup }
\def\title#1{\vskip\frontpageskip \titlestyleb{#1} \vskip\headskip }
%
%

%
%
\paperfootline={\hss\iffrontpage\else\ifp@genum%
                \tenrm --\thinspace\folio\thinspace --\hss\fi\fi}
%
%

%

%
%

%
\def\contr#1#2#3{\vbox{\ialign{##\crcr
          \hskip #2pt\vrule depth 4pt
          \hrulefill\vrule depth 4pt\hskip #3pt
          \crcr\noalign{\kern-1pt\vskip0.125cm\nointerlineskip}
          $\hfil\displaystyle{#1}\hfil$\crcr}}}
\def\leftrightarrowfill{$\m@th\mathord-\mkern-6mu%
  \cleaders\hbox{$\mkern-2mu\mathord-\mkern-2mu$}\hfill
  \mkern-6mu\mathord\leftrightarrow$}
\def\overleftrightarrow#1{\vbox{\ialign{##\crcr
      \leftrightarrowfill\crcr\noalign{\kern-\p@\nointerlineskip}
      $\hfil\displaystyle{#1}\hfil$\crcr}}}
%

%
%

%
\def\rbox#1{\vbox{\hrule height.8pt%
                \hbox{\vrule width.8pt\kern5pt
                \vbox{\kern5pt\hbox{#1}\kern5pt}\kern5pt
                \vrule width.8pt}
                \hrule height.8pt}}
%
%
\def\sqr#1#2{{\vcenter{\hrule height.#2pt
      \hbox{\vrule width.#2pt height#1pt \kern#1pt
          \vrule width.#2pt}
      \hrule height.#2pt}}}
\def\overbar#1{\vbox{\ialign{##\crcr
          \hskip 1.5pt\hrulefill\hskip 1.1pt
          \crcr\noalign{\kern-1pt\vskip0.125cm\nointerlineskip}
          $\hfil\displaystyle{#1}\hfil$\crcr}}}
%
%
%

%
%

%
%
\mathchardef\Lag="724C
%
%

%

%
\def\addeqno{\ifnum\equanumber<0 \global\advance\equanumber by -1
    \else \global\advance\equanumber by 1\fi}
\def\undereq#1{\mathop{\vtop{\ialign{##\crcr
      $\hfil\displaystyle{#1}\hfil$
      \crcr\noalign{\kern3\p@\nointerlineskip}
      \crcr\noalign{\kern3\p@}}}}\limits}
\def\overeq#1{\mathop{\vbox{\ialign{##\crcr\noalign{\kern3\p@}
      \crcr\noalign{\kern3\p@\nointerlineskip}
      $\hfil\displaystyle{#1}\hfil$\crcr}}}\limits}
%

%
%

%
%
\def\journal#1&#2(#3){\unskip, {\sl #1}{\bf #2}(19#3)}
\def\andjournal#1&#2(#3){{\sl #1}{\bf #2}(19#3)}
\def\andvol&#1(#2){{\bf #1}(19#2)}

\def\NP{Nucl. Phys. }
\def\PR{Phys. Rev. }

%
%

%
%

%
%

%
\catcode`@=12
%
%

%

%

\catcode`@=11

\newtoks\heth
\newtoks\Heth
\newtoks\rims
\Pubnum={}
\Heth={}
\rims={RIMS-851}
\date={December 1991}
\def\p@bblock{\begingroup \tabskip=\hsize minus \hsize
   \baselineskip=1.5\ht\strutbox \topspace-2\baselineskip
   \halign to\hsize{\strut ##\hfil\tabskip=0pt\crcr
   \the\Pubnum\cr \the\Heth\cr
    \the\rims\cr \the\date\cr }\endgroup}

\catcode`@=12


\def\symbol#1/#2/#3/#4/#5/#6/{
     \left( \matrix{#1 & #2 & #3 \cr #4 & #5 & #6 \cr}  \right) }
\def\rwsymbol#1/#2/#3/#4/#5/#6/{
     \left\{  \matrix{#1 & #2 & #3 \cr #4 & #5 & #6 \cr}
\right\}^{RW} }
\def\wsymbol#1/#2/#3/#4/#5/#6/{
     \left\{  \matrix{#1 & #2 & #3 \cr #4 & #5 & #6 \cr}  \right\}}
\def\qsymbol#1/#2/#3/#4/#5/#6/{
     \left\{  \matrix{#1 & #2 & #3 \cr #4 & #5 & #6 \cr}  \right\}_q}
\def\qwsymbol#1/#2/#3/#4/#5/#6/{
     \left|  \matrix{#1 & #2 & #3 \cr #4 & #5 & #6 \cr}  \right|_q}
\def\cg#1/#2/#3/#4/#5/#6/{ \left( \matrix{#1 & #2 & #3 \cr #4 & #5 & #6
                                                \cr} \right)}

\pubnum={1088}          
\heth={91/13}           


\titlepage

\title{Partition Functions and Topology-Changing Amplitudes
  \break in the 3D  Lattice Gravity of Ponzano and Regge}

\author{Hirosi Ooguri\foot{e-mail addresses : ooguri@jpnrifp.bitnet
 and  ooguri@kekvax.kek.ac.jp }}

\address{Research Institute for Mathematical Sciences \break
         Kyoto University,
         Kyoto 606, Japan}

\abstract{We define a physical Hilbert space for the three-dimensional
lattice gravity of Ponzano and Regge and establish its isomorphism
to the one in the $ISO(3)$ Chern-Simons theory.
It is shown that, for a handlebody of any genus, a Hartle-Hawking-type
wave-function of the lattice gravity transforms into the corresponding
state in the Chern-Simons theory under this isomorphism.
Using the Heegaard splitting
of a three-dimensional manifold, a
partition functions of each of these theories is expressed as
an inner product of such wave-functions.
Since the isomorphism preserves the inner products,
the partition functions of the two theories
are the same for any closed orientable manifold.
We also discuss on a class of topology-changing amplitudes
in the lattice gravity and their relation to the ones
in the Chern-Simons theory.}

\endpage

\sequentialequations

\section{Introduction}

In 1968,
Ponzano and Regge derived the following asymptotic form of the
Racah-Wigner $6j$-symbol for large angular momenta $j_i$'s
\REF\PR{G.~Ponzano and T.~Regge,
in {\it Spectroscopic and Group
Theoretical Methods in Physics}, ed. F.~Block (North-Holland,
Amsterdam, 1968).}[\PR].
$$
\eqalign{
   (-1)^{\sum_{i=1}^6 j_i} \wsymbol j_1/j_2/j_3/j_4/j_5/j_6/
    & \sim {1 \over \sqrt{12 \pi V}}
           \cos (S_{Regge} + \pi/4) \cr
   &~~~~(j_i \in {\bf Z}_{\geq 0}). \cr}
\eqn\asymptotic
$$
To explain the notations in  the right hand side,
it is useful to imagine
a tetrahedron and associate $j_i$'s to its edges as in Fig. 1.
In the following, we call this as coloring of the tetrahedron.
Since the $6j$-symbol has the tetrahedral symmetry,
we can uniquely associate it to the colored
tetrahedron.
Now we regard $(j_i+{1 \over 2})$ as a length of the $i$-th edge
of the tetrahedron.
The factor $V$ in the right hand side of \asymptotic\
is defined as
a volume of such a tetrahedron, and $S_{Regge}$ is given by
$$
   S_{Regge} = \sum_{i=1}^6 \theta_i(j_i+ {1 \over 2}),
\eqn\regge
$$
where $\theta_i$ is the angle between the outward normals of the two
faces separated by the $i$-th edge.

What is remarkable about this formula is that
$S_{Regge}$ is nothing but the Regge action
\REF\Regge{T.~Regge \journal {\sl Nuovo Cimento}
&19 (61) 558.}[\Regge] for the single tetrahedron.
Suppose there is a three-dimensional
manifold $M$ which is decomposed into a collection of tetrahedra.
If we assume that each tetrahedron is filled in with flat space
and the curvature of $M$ is concentrated on the edges of
the tetrahedron, a metric $g_{\mu\nu}$ on $M$ is specified once
the length $(j+{1 \over 2})$ of each edge is fixed. The Einstein-Hilbert
action $\int d^3x \sqrt{g} R$ is then a function of $j$'s on the edges and
it is given by summing
the Regge action \regge\ over all the tetrahedra in $M$.
Thus, as a model for the three-dimensioal Einstein
gravity, Ponzano and Regge considered
a lattice statistical model whose dynamical
variables are the angular
momenta $j$'s on the edges and whose weight is given by
a product of the $6j$-symbols over
all the tetrahedra in $M$ (including the sign-factor $(-1)^{\sum_i j_i}$
in the left hand side of \asymptotic).

In the lattice gravity, we sum over geometries of $M$
based on its simplicial decomposition.
In one approach, size and shape of each simplex are fixed,
and the quantum flctuation of the geometry is evaluated
by summing over all the possible ways of gluing the simplices
together. The recent studies on two-dimensional gravity
are mostly based on this approach\REF\twod{V.A.~Kazakov \journal
Mod. Phys. Lett. &A4 (89) 2125; ~~~~~~~~~~~~~~~~~~\break
M.R.~Douglas and S.H.~Shenker \journal Nucle. Phys.
&B335 (90) 635; ~~~~~~~~~~~~~~~\break
D.J.~Gross and A.A.~Migdal \journal Phys. Rev. Lett.
&64 (90) 127; ~~~~~~~~~~~~~\break E.~Brezin and
V.A.~Kazakov \journal Phys. Lett. &236B (90) 144.} [\twod].
In the other approach, one fixes the lattice structure
and sums over the lattice lengths [\Regge].
The lattice model of Ponzano and Regge belongs to
the latter approach.

In both of these approaches, it is important to know if the
lattice model has a nice continuum limit. In this respect,
it has already been pointed out by Ponzano and Regge
that their lattice model can be made scale-invariant
with appropriate modification of the statistical weight.
Let us take the tetrahedron in Fig. 1, and decompose it
into four small tetrahedra as in Fig. 2. There are
four edges inside of the original tetrahedron, and we
put angular momenta $l_1,...,l_4$ on them.
Corresponding to the four tetrahedra, we consider
the following product of the $6j$-symbols.
$$
  (-1)^{\sum_i l_i}
  \wsymbol j_1/j_2/j_3/l_1/l_2/l_3/
  \wsymbol j_4/j_6/j_2/l_3/l_1/l_4/
  \wsymbol j_3/j_4/j_5/l_4/l_2/l_1/
  \wsymbol j_1/j_5/j_6/l_4/l_3/l_2/
\eqn\summand
$$
Now we are going to sum this weight over the coloring $l_i$ on
the internal edges. The summation can be performed analytically if
we multiply an additional factor $\prod_{i=1}^4 (2l_i+1)$
to the summand \summand.
By using the identity due to Biedenharn and Elliot, the summation
over $l_1$ can be done as
$$
\eqalign{  &\sum_{l_1} (2l_1+1)
  (-1)^{\sum_i l_i}
  \wsymbol j_4/j_6/j_2/l_3/l_1/l_4/
  \wsymbol j_3/j_4/j_5/l_4/l_2/l_1/
  \wsymbol j_1/j_5/j_6/l_4/l_3/l_2/ \cr
   & ~~~~~=(-1)^{\sum_i j_i}
   \wsymbol j_1/j_2/j_3/j_4/j_5/j_6/
  \wsymbol j_1/j_5/j_6/l_4/l_3/l_2/ . \cr}
$$
We can then sum over $l_4$ using the orthonormality
of the $6j$-symbol
$$
  \sum_{l_4} (2l_4+1)
  \wsymbol j_1/j_5/j_6/l_4/l_3/l_2/^2
    = {1 \over 2j_1 + 1}.
$$
Thus we are left with the sum over $l_2$ and $l_3$ as
$$
\eqalign{& ~~~~~~~  \sum_{l_1,...,l_4}    (-1)^{\sum_i l_i}
  (2l_1+1)(2l_2+1)(2l_3+1)(2l_4+1) \cr
  &~~~~~~~~~~~~~\times \wsymbol j_1/j_2/j_3/l_1/l_2/l_3/
  \wsymbol j_4/j_6/j_2/l_3/l_1/l_4/
  \wsymbol j_3/j_4/j_5/l_4/l_2/l_1/
  \wsymbol j_1/j_5/j_6/l_4/l_3/l_2/ \cr
  &~~~~~=
    (-1)^{\sum_i j_i}
   \wsymbol j_1/j_2/j_3/j_4/j_5/j_6/ ~~{1 \over 2j_1 +1}
    \sum_{|l_2-l_3| \leq j_1 \leq l_2+l_3}
       (2l_2+1)(2l_3+1) .\cr}
\eqn\renormalize
$$
However, the sum over $l_2$ and $l_3$ in the right hand
side is divergent. In order to regularize it, we cut off the
summation by $l_i \leq L$ and rescale the summand of \renormalize\ by
multiplying a factor $\Lambda(L)^{-1}$, where
$$
  \Lambda(L) = {1 \over 2j_1 + 1}
         \sum_{ l_2,l_3 \leq L \atop |l_2-l_3| \leq j_1 \leq l_2+l_3}
       (2l_2+1)(2l_3+1) .
\eqn\lambdadef
$$
For a sufficiently large value of $L$, $\Lambda(L)$ becomes independ on
$l_1$ and behaves as $\Lambda(L) \sim 4L^3/3$ for $L \rightarrow \infty$.
After multiplying this factor, we can take $L$ to $\infty$
and the divergence is removed.
Thus, with the additional factor
$\Lambda(L)^{-1} \prod_i (2l_i+1)$, the sum of
\summand\ over the coloring $l_i$ on the internal edges in Fig. 2
reproduces the weight
$$
(-1)^{\sum_i j_i} \wsymbol j_1/j_2/j_3/j_4/j_5/j_6/
$$
for the original tetrahedron in Fig. 1.

Based on this observation,
Ponzano and Regge defined a partition function $Z_M$ for
the manifold $M$ by
$$
 Z_M = \lim_{L \rightarrow \infty}
    \sum_{\{ j: j \leq L \}} \prod_{vertices} \Lambda(L)^{-1}
                     \prod_{edges} (2j+1)
                     \prod_{tetrahedra} (-1)^{\sum_i j_i}
      \wsymbol j_1/j_2/j_3/j_4/j_5/j_6/.
\eqn\partition
$$
Due to the identity \renormalize,
$Z_M$ is invariant under the refinement of any
tetrahedron in $M$ into four smaller tetrahedra.
Namely this lattice model
is at a fixed point the renormalization group
transformation\foot{Although the divergence
due to the scale-invariance of the model is regularized in \partition\
by multiplying the factor $\Lambda(L)^{-1}$, it is not obvious that
$Z_M$ defined in the above is finite. We will examine this point
in Section 3.}.

Because of this property, one may suspect that the lattice gravity
of Ponzano and Regge can be related to some quantum field theory
in the continuum. Although there have been some works on physical
interpretation of this model\REF\perry{B.~Hasslacher and M.J.~Perry
\journal Phys. Lett. &103B (81) 21.} [\perry], little progress had been
made on the continuum limit of this model until recently.
Last year, Turaev and Viro\REF\TV{V.G.~Turaev and
O.Y.~Viro, ``{\it State Sum  Invariants of 3-Manifolds
and Quantum 6j-Symbols},'' preprint (1990).} studied
the $q$-analogue of the Ponzano-Regge model,
and found that its partition function is invariant
under a class of transformations
larger than the renormalization group in the above.
Moreover they have shown
that any two tetrahedral decompositions of
$M$ can be related by a sequence of such transformations.
Therefore the partition is independent of the tetrahedral decomposition
and depends only on the topology of $M$.
Although they have studied
the $q$-analogue, their argument is directly applicable to
the original model of Ponzano and Regge.
Thus it is natural to expect that
the model of Ponzano and Regge and its $q$-analogue by
Turaev and Viro are equivalent to
some topological field theories. Indeed, in the paper [\TV],
Turaev and Viro have conjectured that the partition function
of their $q$-analogue model is equal to the absolute value square
of the partition function of the $SU(2)$ Chern-Simons theory
\REF\CS{E.~Witten
\journal Commun. Math. Phys. &121 (89) 351.}[\CS]
of level $k$ ($q = e^{2\pi i/(k+2)}$)
when the manifold $M$ is orientable.

In the previous paper\REF\OS{H.~Ooguri
and N.~Sasakura \journal {\sl Mod. Phys. Lett}
&A6 (91) 3591.} [\OS],
the author and Sasakura have examined physical
states in the lattice gravity of Ponzano and Regge and suggested
that they are related to physical states of the
$ISO(3)$ Chern-Simons theory whose action is given by
$$
   S_{CS}(e,\omega) = \int d^3x e^a \wedge (d \omega^a
                            + \epsilon^{abc} \omega^b \wedge
                                                \omega^c ),
\eqn\csaction
$$
where $e^a$ and $\omega^a$ ($a = 1,2,3$)
are one-forms on $M$ with adjoint indices of $SO(3)$.
If we identify them as a dreibein and a spin-connection
following the observation by Witten\REF\W{E.~Witten \journal
Nucl. Phys. &B311 (88/898) 46.}, the action $S_{CS}$ may be
regarded as the Einstein-Hilbert action $\int e \wedge R$
in the first order formalism.

In this paper, we extend the analysis of [\OS] and show that
the partition function of the lattice gravity of Ponzano
and Regge agrees with the one
of the $ISO(3)$ Chern-Simons theory for any orientable manifold.
This result corresponds to the $k \rightarrow \infty$ limit of
the cojecture by Turaev and Viro.
In Section 2, we
define a physical Hilbert space for the lattice gravity and establish
its isomorphism to the physical Hilbert space of
the Chern-Simons theory. We show later in Section 4 that
this isomorphism preserves the inner products of the two Hilbert spaces.
In Section 3, we compute the
Hartle-Hawking-type wave-functions of the lattice gravity
for a handlebody of any genus
and show that it transformes into the corresponding state
in the Chern-Simons
theory under the isomorphism. By
gluing Hartle-Hawking-type wave-functions, one can compute
a partition function for any closed orientable manifold. We check in
Section 4 that this gluing procedure is compatible with
the isomorphism. Therefore the partition functions of these
two theories are the same, as far as they are finite.
We also study some class
of topology-changing amplitudes in the lattice gravity
and their relation to the ones in the Chern-Simons
theory. In the last section, we discuss on interpretations of
these results and their extensions.

In the course of this work, the author was informed of a paper
by Turaev\REF\Turaev{V.G.~Turaev, ``{\it Quantum Invariants of
3-Manifolds and A Glimpse of Shadow Topology},''
hand-written manuscript (1991).} [\Turaev]
where he announces to have
proven the equivalence of the $q$-analogue lattice model
and the Chern-Simons theory for finite $k$. Details of his
derivation not being available, it is not clear to the
author how his approach is related to the one presented here.

\section{Wave-Functions}

In the lattice gravity of Ponzano and Regge, one can define a
discretized verion of the Wheeler-DeWitt equation which
characterizes physical states in the theory.
On the other hand, in the $ISO(3)$ Chern-Simons theory,
a physical state is given by a gauge-invariant half-density
$\Phi(\omega)$ on the moduli space of
a flat $SO(3)$ connection $\omega$ on
a two-dimensional surface $\Sigma$. In this section, we establish
a correspondence between physical states in the lattice gravity
and in the continuum Chern-Simons theory. This
correspondence will be used in the later sections to
compare partition functions and topology-changing amplitudes
in those two theories.

First we should clarify what we mean by physical states in the
lattice gravity. To motivate our definition of physical states,
let us consider a closed three-dimensional manifold
$M$ and decompose it into three parts, $M_1$, $M_2$ and $N$,
as in Fig. 3,
where $N$ has a topology of $\Sigma \times [0,1]$
with $\Sigma$ being a closed orientable
two-dimensional surface, and $M_i$ ($i=1,2$)
has a boundary which is isomorphic to $\Sigma$. The manifold
$M$ is reconstructed by gluing the boundaries of $N$ with
$\partial M_1$ and $\partial M_2$.

Corresponding to this decomposition of $M$, the partition
function $Z_M$ of the manifold $M$ can be expressed
as a sum of products of three components each of which
is associated to $M_1$, $M_2$ and $N$.
To find such an expression,
we note that the partition function
$Z_M$ is independent of a choice of
tetrahedral decomposition of $M$. Therefore
we can place tetrahedra in $M$ in such a way that
$M_1$, $M_2$ and $N$ do not share a tetrahedron, namely
their boundaries are triangulated by the
faces of the tetrahedra.
Corresponding to this tetrahedral decomposition,
we can express $Z_M$ as
$$
 Z_M = \sum_{c_1 \in C(\Delta_1) \atop c_2 \in C(\Delta_2)}
           Z_{M_1,\Delta_1}(c_1)
     \Lambda^{-n(\Delta_1)}
P_{\Delta_1,\Delta_2}(c_1,c_2) \Lambda^{-n(\Delta_2)}
           Z_{M_2,\Delta_2}(c_2),
\eqn\decomp
$$
Here $\Delta_i$ ($i=1,2$) denotes the triangulations of
the boundary $\partial M_i$,
$C(\Delta_i)$ is a set of all the possible colorings on
$\Delta_i$, and $n(\Delta_i)$ is a number of vertices on $\Delta_i$\foot{
For conciseness of equations, here and in the following, we do not
write the cut-off parameter $L$ explicitly.}.
The factor $Z_{M_i, \Delta_i}(c_i)$ is given by the sum over
all the possible coloring on the edges interior of $M_i$
$$
 \eqalign{ Z_{M_i, \Delta}(c_i) = & \prod_{edges~on~\Delta_i}
                        (-1)^{2j}\sqrt{2j+1} \cr
  & \times \sum_{coloring}
            \prod_{vertices \atop interior~of~M_i}
                            \Lambda^{-1}
      \prod_{edges \atop interior~of~M_i}
                 (2j+1) \cr
  &~~~~~~~~~\times
               \prod_{tetrahedra~in~M_i}
      (-1)^{\sum_i j_i}
     \wsymbol j_1/j_2/j_3/j_4/j_5/j_6/ , \cr}
$$
where we keep fixed the coloring $c_i$ on the edges on
$\partial M_i$ (Fig. 4).
Similarly $P_{\Delta_1,\Delta_2}(c_1,c_2)$
is given by a sum over all the possible colorings on the interior edges
of $N$ with
fixed colorings $c_1$ and $c_2$ on $\partial N \simeq \Sigma + \Sigma$.

\def\proj#1/#2{P_{\Delta_{#1},\Delta_{#2}}}

Since $\proj 1/2$ is independent of the tetrahedral decomposition
of the interior of $N$, it satisfies the following remarkable
property,
$$
   \sum_{c_2 \in C(\Delta_2)} \proj 1/2 (c_1,c_2)
      \Lambda^{-n(\Delta_2)}
           \proj 2/3 (c_2,c_3) = \proj 1/3 (c_1,c_3) .
\eqn\projection
$$
Therefore we can
define an operator ${\cal P}$
$$
     {\cal P}[\phi_\Delta](c) =
          \sum_{c' \in C(\Delta)} P_{\Delta,\Delta}(c,c')
                \Lambda^{-n(\Delta)} \phi_\Delta(c') ,
$$
which acts as a projection operator (${\cal P} \cdot {\cal P} =
{\cal P}$) on
a space of functions on $C(\Delta)$.
By using \projection, we can rewrite \decomp\ as
$$
  Z_M = \sum_{c_1, c_2}
         {\cal P}[Z_{M_1,\Delta_1}](c_1')
     \Lambda^{-n(\Delta_1)} \proj 1/2 (c_1,c_2)
     \Lambda^{-n(\Delta_2)}
  {\cal P}[Z_{M_2,\Delta_2}](c_2').
$$
One sees that ``states'' propagating from $M_1$ to $M_2$ through
$N$ are projected out by ${\cal P}$.
Thus it is natural to define a physical Hilbert space $H(\Delta)$
for the triangulated surface $\Sigma$
as a subspace projected out by ${\cal P}$,
i.e.
$$
   \phi_\Delta(c) \in H(\Delta)
     ~~~~~   \Longleftrightarrow ~~~~~\phi_\Delta = {\cal P}[\phi_\Delta]
\eqn\WdW
$$
Since $P_{\Delta,\Delta}$ is associated to the topology $\Sigma \times [0,1]$,
we may regard it as a time evolution operator in the lattice gravity.
Therefore it
should be appropriate to call the physical state condition \WdW\ as a
discretized version of the Wheeler-DeWitt equation\REF\WW{B.S.~DeWitt \journal
{\sl Phys. Rev.} &160 (67) 1113;~~~~~~~~~~~~~~~~~~~~~~~~~~~~~~~~~~
 \break J.A.~Wheeler, in
{\it Batelle Recontres,} eds C.M.~DeWitt and J.A.~Wheeler
(W.A.~Benjamin, 1968).}.
We define an inner product in $H(\Delta)$ by
$$
( \phi_\Delta , \phi_\Delta' )
     = \sum_{c,c' \in C(\Delta)} \phi_\Delta(c)
     \Lambda^{-n(\Delta)}
   P_{\Delta,\Delta}(c,c')  \Lambda^{-n(\Delta)}
                                 \phi_\Delta'(c') .
\eqn\inner
$$
It is easy to see that $Z_{M_1,\Delta}(c)$ and $Z_{M_2,\Delta}(c)$
are real solutions to the
Wheeler-DeWitt equation \WdW\ and the partition function
$Z_{M}$ is given by their inner product
$$
  Z_M = ( Z_{M_1,\Delta} , Z_{M_2,\Delta} ).
\eqn\overlap
$$

Although this definition of $H(\Delta)$ depends of the triangulation
$\Delta$ of $\Sigma$, there is a natural isomorphism
given by the map $\proj 1/2$ between $H(\Delta_1)$
and $H(\Delta_2)$ for any two triangulations $\Delta_1$, $\Delta_2$.
Due to the equation \projection , the map $P_{\Delta_1,\Delta_2}$
preserves the inner product defined by \inner .
It also follows from \projection\ that
the map $P_{\Delta_1,\Delta_2}$
has an inverse and it is given by
$P_{\Delta_2,\Delta_1}$.
Thus we may choose an arbitrary triangulation
in defining the physical Hilbert space for $\Sigma$.

On the other hand, the physical Hilbert space $H_{CS}$
of the $ISO(3)$ Chern-Simons theory consists of
half-densities on the moduli space of a flat $SO(3)$ connection
on $\Sigma$. To see this, we consider the topology
$N = \Sigma \times [0,1]$ again, and decompose the
dreibein $e^a$ and the spin-connection $\omega^a$ ($a=1,2,3$) as
$$
  \eqalign{ e^a = \sum_{i=1,2}e_i^a dx^i &+ e_0^a dt~,~~~
            \omega^a = \sum_{i=1,2} \omega_i^a dx^i + \omega_0^a dt \cr
            &(x^1, x^2) \in \Sigma, ~~t \in [0,1]. \cr}
$$
Corresponding to this decomposition, the Chern-Simons action
\csaction\ takes the form,
$$
    S_{CS}(e,\omega) = \int dt d^2x
                        \epsilon^{ij}
               ( e_j^a \partial_t \omega_i^a + e_0^a F_{ij}^a
                          - \omega_0^a D_i e_j^a ),
$$
where $D_i$ is a covariant derivative given by $\omega_i^a$
and $F_{ij}^a$ is its curvature. From this expression,
one sees that $(\omega_i, \epsilon^{ij}e_j)$ are cannonically
cojugate to each other, while $e_0$ and $\omega_0$ are
Lagrange multipliers and impose constraints,
$F_{ij}=0$ and $ D_i e_j - D_j e_i = 0$.
Thus a wave-function of the theory can be represented
by a function $\Phi(\omega)$ of $\omega_i$, a $SO(3)$
connection on $\Sigma$. The constraint $F_{ij}=0$
implies that $\Phi(\omega)$ should vanish unless
$\omega$ is flat, and $\epsilon^{ij} D_i e_j \Phi(\omega)
 = i D_i {\delta \over \delta \omega_i} \Phi(\omega) =0$
means that $\Phi(\omega)$ is invariant under the
gauge-transformation $\omega_i \rightarrow \omega_i + D_i \lambda$.
The inner product in $H_{CS}$ is given by the integral
$$
  ( \Phi_1 , \Phi_2 )_{CS}
    = \int [d \omega ]
          \delta(F_{ij}) \Phi_1^*(\omega)
                                     \Phi_2(\omega).
\eqn\csinner
$$
Thus a physical wave-function is a half-density on the moduli space
of a flat $SO(3)$ connection.

Now we would like to show that there is a natural isomorphism
between $H(\Delta)$ and $H_{CS}$.  To interpolate
between the two Hilbert spaces, we introduce the following
(over-complete) basis for $H_{CS}$ constructed from
Wilson-lines $U_j(x,y)$ ($x,y \in \Sigma, j=0,1,2,...$),
$$
    U_j(x,y) = P \exp (\int_x^y \omega^a t^a_j ),
$$
where $P \exp $ denotes the path-ordered exponential and
$t_j^a$ ($a=1,2,3$) is the spin-$j$ generator of $SO(3)$.
Under a gauge transformation $\omega \rightarrow
\Omega^{-1} \omega \Omega + \Omega^{-1} d \Omega$,
the Wilson-line behaves as $U(x,y) \rightarrow
\Omega(x)^{-1} U(x,y) \Omega(y)$. Now consider their
tensor product $\otimes_i U_{j_i}(x_i, y_i)$. To
make this gauge-invariant, we need to contract group
indices of $U_j$'s so that the gauge factor $\Omega$
cancels out. In the case of the group $SO(3)$,
invariant tensors we can use to contract
group indices are the Clebsch-Gordan coefficient
$\langle j_1 j_2 m_1 m_2 | j_3 m_3 \rangle$ and the
metric
$$
\eqalign{ g_{mm'}^{j} & = (-1)^{j-m} {1 \over \sqrt{2j+1}}
                            \delta_{m+m',0} \cr
          g_{j}^{mm'} & = (-1)^{j+m} \sqrt{2j+1} \delta_{m+m',0}.\cr}
$$
Actually it is
more convenient to use the cyclic-symmetric $3j$-symbol
given by
$$
  \cg j_1/j_2/j_3/m_1/m_2/m_3/ = {(-1)^{j_1-j_2-m_3} \over
                              \sqrt{2j_3+1}}
           \langle j_1 j_2 m_1 m_2 | j_3 -m_3 \rangle,
$$
rather than the Clebsch-Gordan coefficient.
We regard $m_i$'s in the $3j$-symbol as lower indices
which can be raised by the metric $g_{j_i}^{m_im_i'}$.
When three Wilson-lines meet together
at the same point on $\Sigma$, we can use
the $3j$-symbol and the metric $g_{mm'}^{j}$ to contract
their group indices.
We can also connect two
Wilson-lines by the metric if they carry the same spin.
The gauge-invariant function constructed this way
corresponds to a colored trivalent graph $Y$ on $\Sigma$,
where a contour from $x$ to $y$ in $Y$ with color $j$
corresponds to a Wilson-line $U_j(x,y)$, and a three-point
vertex in $Y$ represents the $3j$-symbol\foot{
There may be a pair of Wilson-lines intersecting with each
other, which cannot be described as a part of a trivalent graph
as it is. In such a case, we may cut the Wilson-lines
at the intersecting point and use the orthonormality of the $3j$-symbols,
$$
 \delta_{m_1,m_1'} \delta_{m_2,m_2'} =
\sum_{j_3,m_3} (2j_3+1)
\cg j_1/j_2/j_3/m_1/m_2/m_3/
\cg j_1/j_2/j_3/m_1'/m_2'/m_3/
$$
to replace the intersection by two vertices and an
infinitesimal Wilson-line connecting the vertices.}.
Due to the cyclic symmetry of the $3j$-symbol,
to each graph $Y$ on the orientable surface $\Sigma$,
we can associate such a
gauge-invariant function uniquely.

A physical wave-function of the Chern-Simons
theory is obtained from such a network of Wilson-lines
by restricting the support of the function on flat $SO(3)$ connections.
This restriction however gives rise to linear dependence among the
Wilson-line networks.
Specifically, if two graphes $Y$ and $Y'$ are homotopic, the corresponding
gauge-invariant functions have the same value on a
flat connection. Since there is one to one correspondence between
a homotopy class of colored trivalent graphes on $\Sigma$
and a triangulation of $\Sigma$ with coloring on their sides,
we may parametrize the gauge-invariant function by a colored
triangulation defined by a pair $(\Delta, c)$ ($c \in C(\Delta)$)
rather than a trivalent graph $Y$.
In this way, to each colored triangulation, we can associate
a physical wave-function $\Psi_{\Delta,c}$
of the Chern-Simons theory.
An arbitrary wave-function $\Phi(\omega)$ is expanded
in terms of them as
$$
\Phi(\omega) = \sum_\Delta \sum_{c \in \Delta} \varphi_\Delta (c)
                \Lambda^{-n(\Delta)} \Psi_{\Delta,c}(\omega).
\eqn\expansion
$$

Now we are in a position to establish a correspondence between
a solution to the discretized Wheeler-DeWitt equation \WdW\
and a physical state in the Chern-Simons theory.
To understand the correspondence, the following fact is
most important. When evaluated on a flat connection $\omega$,
$\Psi_{\Delta,c}(\omega)$ are not yet linearly independent,
but they obey the following relations,
$$
 \Psi_{\Delta,c}(\omega) = \sum_{c' \in C(\Delta')} P_{\Delta,\Delta'}(c,c')
               \Lambda^{-n(\Delta')}
      \Psi_{\Delta',c'} (\omega).
\eqn\recomb
$$
Furthermore they are the only linear relations on
$\Psi_{\Delta,c}$.

Before proving \recomb , let us
examine its consequences.
By substituting \recomb\ into \expansion,
we obtain
$$
  \Phi(\omega) = \sum_{c \in C(\Delta)} \phi_\Delta (c)
       \Lambda^{-n(\Delta)}  \Psi_{\Delta,c}(\omega),
\eqn\isomorphism
$$
where $\phi_\Delta(c)$ is defined by
$$
   \phi_\Delta(c) = \sum_{\Delta'} \sum_{c' \in C(\Delta')}
                    P_{\Delta,\Delta'}(c,c')
 \Lambda^{-n(\Delta')}         \varphi_{\Delta'}(c')
$$
for an arbitrary {\it fixed} triangulation $\Delta$ of $\Sigma$.
It follows from \projection\ that $\phi_\Delta(c)$
solves the Wheeler-DeWitt equation \WdW\ of the lattice gravity.
$$
   \phi_\Delta = {\cal P}[\phi_\Delta].
$$
Thus, to each solution $\phi_\Delta(c)$ of the
Wheeler-DeWitt equation, there is a physical state
$\Phi(\omega)$ of the Chern-Simons theory given by \isomorphism.
Since \recomb\ are the only relations among $\Psi_{\Delta,c}$'s,
this correspondence between $\phi_\Delta$ and $\Phi$ is one to one.
In Section 4, we will show that the inner product of Wilson-line
networks $( \Psi_{\Delta_1,c_1}, \Psi_{\Delta_2,c_2} )_{CS}$
in the Chern-Simons theory
is equal to $P_{\Delta_1,\Delta_2}(c_1,c_2)$ for the lattice
gravity, upto a constant factor.
Therefore the map from $H(\Delta)$ to
$H_{CS}$ defined by \isomorphism\ preserves their inner
products.
Thus \isomorphism\ gives
the isomorphism between the physical Hilbert spaces of the lattice
gravity and the Chern-Simons theory.

Now we would like to prove that the relations \recomb\ indeed hold,
and that they are the only relations among $\Psi_{\Delta,c}$'s.
We will show this by mathematical induction with respect to the number
of tetrahedra in $N = \Sigma \times [0,1]$.
When the number is zero, the triangulations $\Delta$ and $\Delta'$
must be identical
and they are attached to each other. In this case,
\recomb\ is an obvious identity. Now we are going to pile tetrahedra
one on another and increase the number of tetrahedra in $N$.
Since one tetrahedra has four faces, there are
three ways to attach one on another.

\noindent
(i) Choose one of the faces of the tetrahedron and attach it to one of
the triangles on the surface $\Sigma$ of $N$ (Fig. 5).

\noindent
(ii) Attach two faces of the tetrahedron to two neighbouring triangles
on $\Sigma$ (Fig. 6).

\noindent
(iii) Attach three faces of the tetrahedron
to three neighbouring triangles on $\Sigma$
(figure obtained by inverting the arrows in Fig. 5).

Let us first check that the induction holds in the second move
in the above list. Consider a part of the Wilson-line network
of $\Psi_{\Delta,c}$ which looks like the diagram in the
right hand side of Fig. 6.
Because of the flatness of $\omega$, we can take the Wilson-line
colored by $k$ and
make its length to be arbitrary small without changing the value
of $\Psi_{\Delta,c}(\omega)$. When
its end-points meet with each other, the Wilson-line can
be replace by an identity.
Since the group indices of the Wilson-line $U_k(x,y)$ at the end-points
$x$ and $y$ are
contracted with the $3j$-symbols,
in the limit $x \rightarrow y$ when $U_j(x,y)$ becomes an identity,
the function $\Psi_{\Delta,c}$ should contain a sum of product of these
$3j$-symbols.
Now there is a formula which relate two different ways
of summing $3j$-symbols,
$$
\eqalign{
&   \sum_{mm'}  g_{k}^{mm'}
                \cg j_2/j_3/k/m_2/m_3/m/
                 \cg j_4/j_1/k/m_4/m_1/m'/ \cr
&~~~~~~ = \sum_{l}  (-1)^{j_1+j_2+j_3+j_4}
            \sqrt{(2k+1)(2l+1)}
              \wsymbol j_1/j_2/l/j_3/j_4/k/ \cr
&~~~~~~~~~~~~~~~~~~~~~~~~~\times \sum_{nn'} g_{l}^{nn'}
      \cg j_1/j_2/l/m_1/m_2/n/ \cg j_3/j_4/l/m_3/m_4/n'/
                \cr}
\eqn\cgrelation
$$
The left hand side of this equation corresponds to the diagram in the
left hand side of Fig. 6. These four external Wilson-lines
are recombined in the right hand side; the Wilson-lines of
$j_1$ and $j_2$ make a pair and they are connected to $j_3$ and $j_4$
by an infinitesimal Wilson-line with color-$l$. This
is exactly the right hand side of Fig. 6.
Therefore we obtain
$$
  \Psi_{\Delta, c_{k}} =
  \sum_{l}  (-1)^{j_1+j_2+j_3+j_4} \sqrt{(2k+1)(2l+1)}
              \wsymbol j_1/j_2/l/j_3/j_4/k/
           \Psi_{\widetilde{\Delta}, \widetilde{c}_{l}},
\eqn\recombone
$$
where the triangulation  $(\Delta,c_k)$ contains
two triangles colored as
in the left hand side of Fig. 6, and it is
replaced by
the ones in the dual position
in $(\widetilde{\Delta}, \widetilde{c}_{l})$.

To prove \recomb\ inductively, suppose that we have used $n$-tetrahedra
in constructing the projection operator $P$
for the topology $N = \Sigma \times [0,1]$.
When we add one more tetrahedron to $N$, as is prescribed in (ii),
the corresponding projection operator $P'$
is obtained from $P$
by multiplying to it an appropriate factor involving the $6j$-symbol,
and by summing over coloring on the
common side of two neighbouring triangles to which the new tetrahedron is
attached. This operator $P'$ is obtained exactly by substituting \recombone\
into the right hand side of \recomb. Therefore the inductive proof of
\recomb\ holds when we add one tetradedron in the second move in the list.

We can also add a tetrahedron as in (i) or (iii) in the list.
If the network contains a contractable loop with several
external Wilson-lines attached, by repeatedly using
the identity \cgrelation, the loop can be recombined into
a tree-like diagram with a one-loop tadpole.
The tadpole can be made arbitrarily small, and the infinitesimal
tadpole can be removed by using
$$
  \sum_{mm'} g_{j}^{mm'}
          \cg j/j/J/m/m'/M/
   = \delta_{J,0} \delta_{M,0}.
$$
For example, if the network defined by $(\Delta, c)$
contains a loop with three external lines $j_1$, $j_2$ and
$j_3$ as in the right hand side of Fig. 5,
we man shrink the loop to obtain another
network $(\Delta',c')$ where the three lines meet at one
point. By using the formula,
$$
  \eqalign{ & \sum_{n_{ij}} g_{l_{12}}^{n_{12}n_{12}'}
           g_{l_{23}}^{n_{23}n_{23}'}   g_{l_{31}}^{n_{31}n_{31}'}
     \cg l_{12}/j_2/l_{23}/n_{12}'/m_2/n_{23}/
   \cg l_{23}/j_3/l_{31}/n_{23}'/m_3/n_{31}/
     \cg l_{31}/j_1/l_{12}/n_{31}'/m_1/n_{12}/ \cr
    & ~~~~~= (-1)^{j_1+j_2+j_3} \sqrt{(2l_{12}+1)(2l_{23}+1)(2l_{31}+1)}
          \cr & ~~~~~~~~~~~~~~
{}~~~~~~~~~~~~~~~~~~~~~~~~~~~~~~~~\times
   \wsymbol j_1/j_2/j_3/l_{23}/l_{31}/l_{12}/
      \cg j_1/j_2/j_3/m_1/m_2/m_3/  , \cr}
$$
we can relate the corresponding functions $\Psi_{\Delta,c}$ and
$\Psi_{\Delta',c'}$ as
$$
\Psi_{\Delta,c} = (-1)^{j_1+j_2+j_3}
        \sqrt{(2l_{12}+1)(2l_{23}+1)(2l_{31}+1)}
      \wsymbol j_1/j_2/j_3/l_{23}/l_{31}/l_{12}/
         \Psi_{\Delta',c'},
\eqn\recombtwo
$$
where $l_{ij}$ is the color of the segment of the loop in
$(\Delta,c)$ connecting $j_i$ and $j_j$.
This corresponds to (iii) in the list, and the inductive proof
holds in this move.
The induction for the move (i) is also guaranteed by the same
equation \recombtwo.

In this way, we have proved that the identities \recomb\ holds
forn arbitrary pair of $\Delta$ and $\Delta'$.
Using a variation of the analysis in Appendix D of
\REF\boulatov{B.V.~Boulatov, V.A.~Kazakov,
I.K.~Kostov and A.A.~Migdal, ~~~~~~~~~\break
{\sl Nucl. Phys.} {\bf B273 [FS17]} (1986) 641.}[\boulatov],
one can show that all other relations among $\Psi_{\Delta,c}$
on a flat connection $\omega$ are generated from
\recombone\ and \recombtwo. Therefore \recomb\ are the only
relations among $\Psi_{\Delta,c}$'s.

\section{The Hartle-Hawking-Type Wave-Function}

In the previous section, we defined the isomorphism between
the physical Hilbert spaces of the lattice gravity and the Chern-Simons theory.
In this section, we will show that this isomorphism indeed
identifies wave-functions associated to
the same geometry of the three-dimensional manifold.
The geometry we consider here is a handlebody $M$. To describe $M$,
we embed a closed orientable two-dimensional surface $\Sigma$
into ${\bf R}^3$. The handlebody $M$ is taken as the interior of $\Sigma$.
Associated to such a geometry, we can construct physical states
in both the lattice gravity and the Chern-Simons theory.

In the Chern-Simons theory, the physical wave-function $\Phi_M$
for $M$ is defined as
$$
   \Phi_M(\omega_{|\Sigma}) \delta(F_{ij|\Sigma})~~
= \int_{\omega_{|\Sigma}: fixed}
                [de, d\omega] \exp( i \int_M e \wedge (d\omega + \omega \wedge
                                     \omega) ),
\eqn\hawking
$$
where we perform the functional integral
over $e$ and $\omega$ in the interior of $M$ with a fixed boundary
condition of $\omega$ on $\partial M = \Sigma$. The integration
over $e_{|\Sigma}$ gives rise to $\delta(F_{ij|\Sigma})$ which is explicitly
written in the left hand side of \hawking. This ensures that
the functional integral in the right hand side
gives a physical state of the Chern-Simons
theory. Such a wave-function
may be regarded as a generalization of the Hartle-Hawking
wave-function (The original
wave-function of Hartle
and Hawking \REF\HH{J.~Hartle and S.~Hawking \journal
{\sl Phys. Rev.}& D28 (83) 2960.}[\HH]
corresponds to the case when $\Sigma$ is $S^2$ and the handlebody $M$ is
a three-dimensional ball.).
The wave-function for the lattice gravity is defined in a similar
fashion by fixing a triangulation $\Delta$ and its coloring $c$
of $\Sigma$, and by summing over all possible coloring in the
interior of $M$. This is nothing but $Z_{M,\Delta}(c)$ we have introduced
in Section 2.

In the previous section, we have found that, to each physical state
$\phi_\Delta(c)$ of the lattice gravity, there is a corresponding
state in the Chern-Simons theory defined by \isomorphism.
Therefore it is natural to expect that the wave-functions
$\Phi_M(\omega_{|\Sigma})$ and $Z_{\Delta,M}(c)$ associated to the
same handlebody $M$ are related as
$$
    \Phi_M (\omega_{|\Sigma}) = A_g
                    \sum_{c \in C(\Delta)} Z_{M,\Delta}(c)
       \Lambda^{-n(\Delta)}
                             \Psi_{\Delta,c}(\omega_{|\Sigma}),
\eqn\equivalence
$$
when $\omega_{|\Sigma}$ is a flat connection.
Here $A_g$ is a constant depending only on the genus of the
handlebody $M$.
This indeed is the case as we shall see below.

As was shown by Witten in the case of the Lorentzian Einstein
gravity \REF\Witten{E.~Witten \journal \NP &B323 (89)
113.}[\Witten],
there is a fairly explicity expression for the Hartle-Hawking-type
wave-function $\Phi_M(\omega_{|\Sigma})$.
As well as the constraint $F_{ij|\Sigma}=0$ written explicitly
in \hawking,
the integration over $e$ in \hawking\ imposes that
$\Phi_M$ should vanish unless
$\omega_{|\Sigma}$ have a flat extension $\omega$
interior of the handlebody $M$. This condition can be rephrased as follows.
If the boundary $\Sigma$ of $M$ is of genus $g$, it has $2g$
homology cycles. Among these, there are $g$ cycles which are
contractable in $M$ while other $g$ cycles are not. The
necessary and sufficient condition for $\omega_{|\Sigma}$ to have
a flat extension in $M$ is that its holonomies $U_{(a)}$
($a = 1,...,g$) around these contractable cycles are trivial.
Therefore
$$
    \Phi_M(\omega_{|\Sigma}) = A_g' \prod_{a=1}^g
                              \delta (U_{(a)} - {\bf 1}),
\eqn\holonomy
$$
where $A_g'$ is a constant independent of $\omega_{|\Sigma}$,
and $\delta(U-{\bf 1})$ is a $\delta$-function with respect to
the Haar measure of $SO(3)$.
Thus in order to prove the identity \equivalence, we need to
show that the sum over coloring in the right hand side of the
equation imposes the constraint
$U_{(a)} = {\bf 1}$ on $\omega_{|\Sigma}$.
In the following we will show that,
by recombining the Wilson-lines,
the sum in the right hand side of \equivalence\
reduces to sums over colorings of the contractable cycles as
$$
   \sum_{c \in C(\Delta)} Z_{M,\Delta}(c)
   \Lambda^{-n(\Delta)}
    \Psi_{\Delta,c}(\omega_{|\Sigma})
      = \prod_{a=1}^g \left[ \sum_{j=0}^\infty (2j+1)
                                ~{\sl Tr}(U_{(a)}j)\right].
\eqn\character
$$
The orthonormality and the completeness of the irreducible $SO(3)$
characters \REF\pontryagin{See for example, L.S.~Pontryagin, Chapter 5 of
``{\it Topological Groups},'' \break the second edition, translated by A.~Brown
(Gordon and Breach, 1966).}[\pontryagin] imply
that the right hand side of this equation gives
the product of the $\delta$-functions as in \holonomy.

Now we would like to prove the equation \character.
Suppose we have used $n$-tetrahedra in computing
the wave-function $Z_{\Delta,M}(c)$
for the handlebody $M$. The tetrahedra
must have been placed in such a way
that the boundary $\Sigma$
of $M$ is triangulated as $(\Delta,c)$.
Let us choose one of the tetrahedra attached on the boundary surface.
Since $\omega_{|\Sigma}$
is flat, we can use \recomb\ to remove this tetrahedron,
i.e.
$$
  \sum_{c \in C(\Delta)} Z_{M,\Delta}(c)
  \Lambda^{-n(\Delta)} \Psi_{\Delta,c}(\omega_{|\Sigma})
=  \sum_{c' \in C(\Delta')}
Z_{M,\Delta'}(c)  \Lambda^{-n(\Delta')}
   \Psi_{\Delta',c'}(\omega_{|\Sigma}),
$$
where $\Delta$ is the original triangulation of $\Sigma$ in \character,
and $\Delta'$ is the one which is obtained by removing the tetrahedron
attached on $\Sigma$. In computing $Z_{M,\Delta'}(c)$, the number of
tetrahedra we use is $(n-1)$. By repeating this procedure,
we can eliminate all the tetrahedra in $M$.

To visualize this process,
it is useful to imagine the handlebody $M$ as a balloon whose
surface is of genus $g$. For example, when $\Sigma$ is a torus,
we consider a tube of a tire.
Removing the tetrahedra is then like
reducing the air from the balloon. After gradually
decreasing its volume, the balloon will eventually be flattened.
To describe the flattened balloon, we note that
the surface $\Sigma$ can be constructed from two discs with $g$ holes,
$S_g^+$ and $S_g^-$, by gluing their boundaries together
as shown in Fig. 7.
We call $S_g^+$ and $S_g^-$ as upper and lower parts of $\Sigma$.
The boundaries of the $g$ holes in $S_g^\pm$ correspond to the homology
cycles on $\Sigma$ which are not contractable in $M$.
In the limit when the balloon is flattened, the upper and the lower
parts of $\Sigma$ overlap
one on another. Reflecting the original
tetrahedral decomposition of $M$, $S_g^\pm$ are covered by
triangles. It is not difficult to see that the triangulations of $S_g^+$
and $S_g^-$
must be identical and that they must have the same coloring.
Namely the Wilson-line network in the upper part of $\Sigma$
is the mirror image of the one in the lower part as shown in Fig. 8.

Let us perform the sum over colorings of the Wilson-lines across
the boundaries of $S_g^+$ and $S_g^-$ (for example the Wilson-line
$j_3$ in Fig. 8).
As we did in the previous section, we may take the lengths of
these Wilson-lines arbitrarily small and replace them by ${\bf 1}$.
Because of the reflection symmetry of the
Wilson-lines, we may use the orthonormality of the $3j$-symbols
$$
\eqalign{ &
  \sum_{j_3,m_3,m_3'}
        (-1)^{j_3}\sqrt{2j_3+1}   g_{j_3}^{m_3m_3'}
             \cg j_1/j_2/j_3/m_1/m_2/m_3/
       \cg j_3/j_2/j_1/m_3'/m_2'/m_1'/ \cr
  &~~~~~~~~~~~~~~~~~~~~~= (-1)^{j_1+j_2} \sqrt{(2j_1+1)(2j_2+1)}
 g^{j_1}_{m_1m_1'}  g^{j_2}_{m_2m_2'}  \cr}
\eqn\ortho
$$
to emilinate
the trivalent vertices at the end-points of the Wilson-lines
(as shown in Fig. 9).
Repeating this procedure, we can remove
the vertices on $\Sigma$ one by one.

To understand how the resulting Wilson-line network
looks like, let us examine
the case when $\Sigma$ is a torus, in detail.
In this case, its upper and lower parts are topologically the same as
annuli, and each of them
can be decomposed into two triangles as shown in Fig. 10.
Corresponding to this triangulations,
there are six Wilson-lines
on $\Sigma$ which
are connected by four vertices (Fig. 11a).
We can choose one of the Wilson-lines, say $j_2$ in Fig. 11a,
and remove a pair of vertices at its end-points by using \ortho.
As the result, we obtain a diagram as shown in Fig. 11b.
Because of the flatness of $\omega_{|\Sigma}$, we can move around
the Wilson-line $j_1$ homotopically, and the network in Fig. 11b
can be brought into the one in Fig. 11c.
Now the Wilson-loop consisting of $j_1$ and $j_3$
is contractable
on $\Sigma$, and we end up in Fig. 11d. In this way,
the Wilson-line network on the torus is deformed into a
single Wilson-loop around its homology cycle contractable
in $M$, as shown in Fig. 11e.
Taking into account the weight $Z_{M,\Delta}(c)$,
we have checked
that the resulting summation over $j_4$ reproduces
the right hand side of \character\ for $g=1$.

For $g \geq 2$, we can, for example, choose a triangulation of $S_g^\pm$
as in Fig. 12a. The corresponding Wilson-line network is
shown in Fig. 12b. As in the case of the torus described in the above,
one can follow the deformation of the network and show
that $\sum_c Z_{M,\Delta}(c) \Psi_{\Delta,c}$ for this triangulation
$\Delta$ reduces
to the right hand side of \character.
This proves the identity \equivalence, and we found the factor $A_g$
is equal to $A_g'$ which is related to the normalization of the
path integral \hawking.

\section{Partition Functions and Topology-Changing Amplitudes}

We have found that
the Hartle-Hawking-type wave-functions in
the lattice gravity and the Chern-Simons theory
are related by the isomorphism \expansion\ between the
physical Hilbert spaces of the two theories.
In this section, we will exploit this result
to show that, for any closed orientable manifold $M$,
the partition functions of the two theories
agree with each other.
The idea is to use the Heegard splitting of $M$ \REF\singer{J.~Singer
\journal Transactions of the American Math. Soc. &35 (33) 88}[\singer].
Consider two handlebodies $M_1$ and $M_2$ whose boundaries are of
the same topology $\Sigma$. Since $M_1$ and $M_2$ differ only
by the markings of the homology cycles on their boundaries,
we can glue the boundaries together by their diffeomorphism
and obtain a closed three-dimensional manifold. Moreover it is known that
any closed manifold can be realized in this way. In this construction,
the topology of $M$ is encoded into the topology of $\partial M_1$ and
$\partial M_2$  and how they are glued together.

Corresponding to this splitting of $M$,
the partition function of the
Chern-Simons theory is expressed as an inner product of
the Hartle-Hawking-type wave functions $\Phi_{M_1}$ and $\Phi_{M_2}$,
$$
    Z_M^{(CS)} = ( \Phi_{M_1} , \Phi_{M_2} )_{CS},
\eqn\cspartition
$$
as far as $M$ is orientable.
This formula is derived from the functional
integral expression for $Z_M^{(CS)}$;
the functional integrals over $M_1$ and $M_2$
result in the Hartle-Hawking-type wave-functions
$\Phi_{M_1}$ and $\Phi_{M_2}$, and
the functional integral on the boundary $\partial M_1 \simeq \partial M_2$
corresponds to taking their inner product.
On the other hand, the partition function for the lattice gravity
has also the expression
$$
 Z_M = ( Z_{M_1,\Delta} , Z_{M_2, \Delta} ),
\eqn\prpartition
$$
as we saw in Section 2.
Since the Hartle-Hawking-type wave-functions in the Chern-Simons theory
and the lattice gravity are
related by \equivalence , $Z_M^{(CS)}$ and $Z_M$ are
the same provided the isomorphism \expansion\
preserves the inner products in the two Hilbert spaces,
$H_{CS}$ and $H(\Delta)$.
Thus, in order to establish the equivalence
$Z_M^{(CS)} = Z_M$, we want to show
$$
  ( \Psi_{\Delta_1,c_1} , \Psi_{\Delta_2, c_2})_{CS}
   = A_g^{2} \cdot     P_{\Delta_1,\Delta_2}(c_1,c_2)
\eqn\innerone
$$
or equivalently
$$
\eqalign{&
   \sum_{c_1 \in C(\Delta_1)
           \atop c_2 \in C(\Delta_2)} \Psi_{\Delta_1, c_1}(\omega_1)
                    \Lambda^{-n(\Delta_1)} P_{\Delta_1,\Delta_2}(c_1,c_2)
          \Lambda^{-n(\Delta_2)}
            \Psi_{\Delta_2, c_2}(\omega_2) \cr
       &~~~~~~~~~~~~~~~~~~~~~~~~~~~~~~
       = A_g^{-2} \cdot K(\omega_1,\omega_2) , \cr}
\eqn\innertwo
$$
where $ K(\omega_1, \omega_2)$ is
a kernel for the inner product
$$
    ( \Phi, \Phi')_{CS} = \int [ d \omega_1 ] \delta(F_{1,ij})
 \int [ d \omega_2 ] \delta(F_{2,ij})
    \Phi(\omega_1) K(\omega_1,\omega_2) \Phi(\omega_2),
$$
and it is given in term of the functional integral
$$
    K(\omega_1, \omega_2)
         \delta(F_{1,ij}) \delta(F_{2,ij}) =
    \int_{\omega(t=0)=\omega_1
            \atop \omega(t=1)=\omega_2}
          [ de, d \omega ] \exp( i S_{CS}(e,\omega))
\eqn\kernel
$$
for the topology $N = \Sigma \times [0,1]$.

Now we are going to show that the left hand side of \innertwo\
is proportional to the right hand side. The factor
$A_g^{-2}$ will be fixed later. Since $\Psi_{\Delta_i,c_i}$
is evaluated on a flat connection $\omega_i$,
we may use \recomb\ to rewrite the left hand side
of \innertwo\ as
$$
    \sum_{c \in C(\Delta)} \Lambda^{-n(\Delta)}
           \Psi_{\Delta,c}(\omega_1) \Psi_{\Delta,c}(\omega_2).
\eqn\innerthree
$$
On the other hand, it follows from the functional
integral expression \kernel\ that the kernel $K(\omega_1,\omega_2)$
vanishes unless $\omega_1$ and $\omega_2$
has a flat extension in $N$.
For $N= \Sigma \times [0,1]$,
the flat extension exists if and only if
$\omega_1$ and $\omega_2$ are gauge-equivalent.
Thus we need to show that
the sum over coloring in \innerthree\ imposes the
the constraint, $\omega_1 \simeq \omega_2$

Let us study the case when $\Sigma$ is a torus, in detail.
In this case,
the surface $\Sigma$ can be decomposed into two triangles as shown
in Fig. 13a. The corresponding network of Wilson-lines
is shown in Fig. 13b. A flat connection
$\omega$ on the torus can be specified by holonomies $U$ and
$V$ around the two homology cycles on $\Sigma$.
The wave-function $\Psi_{\Delta,c}$ for the network can
then be regarded as a function of $U$ and $V$. In
the network in Fig. 13b,
the Wilson-line $j_3$ can be made arbitrarily short using
the flatness of $\omega$
and be replaced by an identity. In this case, the wave-function
$\Psi_{\Delta,c}$ is expressed as a function of $U$ and $V$ as
$$
\eqalign{ &
     \Psi_{\Delta;c}(U,V)
           =
      \sum_{m_i,m_i',m_i''}
    U_{j_1 m_1'}^{~m_1}
    V_{j_2 m_2'}^{~m_2} \cr
   &~~~~~~~~~~~~~~~~\times
    g_{j_1}^{m_1' m_1''}
    g_{j_2}^{m_2' m_2''} g_{j_3}^{m_3 m_3''}
            \cg j_1/j_2/j_3/m_1/m_2/m_3/
             \cg j_1/j_2/j_3/m_1''/m_2''/m_3''/
         . \cr}
\eqn\torusnet
$$
Here we marked the homology cycles
on $\Sigma$ in such a way that
the Wilson-lines $j_1$ and $j_2$ wind around cycles
corresponding to the holonomies $U$ and $V$.
The holonomies $U$ and $V$
commute with each other, so they can be
diagonalized simultaneously.
Since the wave-function $\Psi_{\Delta,c}$
is invariant under the simultaneous conjugation,
$U \rightarrow \Omega^{-1} U \Omega$,  $V
\rightarrow \Omega^{-1} V \Omega$, we can
substitute diagonal matrices
$U_{j m'}^{~m} = e^{i m \theta} \delta_{m'}^m$
and $V_{j m'}^{_m} = e^{i m \varphi} \delta_{m'}^m$.
into $U$ and $V$ in \torusnet .

Now we would like to perform the summation,
$$
            \sum_{j_1,j_2,j_3} \Lambda^{-1}
       \Psi_{\Delta,c}(\theta_1,\varphi_1)
       \Psi_{\Delta,c}(\theta_2,\varphi_2),
\eqn\sumtorus
$$
where $(\theta_1,\varphi_1)$ and $(\theta_2,\varphi_2)$
are phases of the holonomies $(U_1,V_1)$ and $(U_2,V_2)$
for $\omega_1$ and $\omega_2$.
Although it is possible to do the summation
for generic values of the phases,
it is more instructive to study
the cases when two among the four phases vanish.
Actually it is enough to study these cases as we shall see below.

Let us consider the case when $V_1=V_2={\bf 1}$.
In this case, the wave-function $\Psi_{\Delta,c}(U_i,V_i)$
is simplified as
$$
\Psi_{\Delta,c}(U_i,V_i={\bf 1})
  = \sqrt{{(2j_1+1)(2j_2+1)(2j_3+1) \over 2j_1 +1}}
         {\sl Tr}(U_{ij_1}).
$$
The summation \sumtorus\ is then performed as
$$
\eqalign{ & \sum_{j_1,j_2,j_3} \Psi_{\Delta,c}(U_1,V_1={\bf 1})
                               \Psi_{\Delta,c}(U_2,V_2={\bf 1}) \cr
          & = \sum_{j_1} {\sl Tr}(U_{1 j_1}) {\sl Tr}(U_{2 j_1})
             \cdot \Lambda^{-1} \cdot
               {1 \over 2j_1+1} \sum_{|j_2-j_3| \leq j_1
                       \leq j_2+j_3} (2j_2+1)(2j_3+1) \cr
          & = \sum_{j_1} {\sl Tr}(U_{1 j_1}) {\sl Tr}(U_{2 j_1})
            = \delta(U_1-U_2). \cr}
$$
Here we have used the definition \lambdadef\ of $\Lambda$
and the orthonormality of the irreducible characters
${\sl Tr}(U_{j_1})$.  Thus
the sum over the coloring in
the left hand side of \innertwo\ indeed imposes
the constraint $U_1 = U_2$ when $V_1=V_2$.
It is straightforward to do the computations in other cases when
$U_2=V_2={\bf 1}$ or $V_1=U_2={\bf 1}$, and we have found that
the sum in \sumtorus\ imposes $U_1=U_2$
and $V_1=V_2$ in both of these cases.

We have seen that the left hand side of \innertwo\
is proportional to $K(\omega_1,\omega_2)$ as far as
two among the four phases are equal to zero.
Let us relax this condition and suppose that
they are not necessarily zero, but
their ratios $\theta_i/\varphi_i$ ($i=1,2$)
are rational numbers. Since \sumtorus\ is invariant
under the modular transformation of $\Sigma$,
we can change the basis of the homology cycles
in such a way that two among the four phases
around the cycles become equal to zero.
The summation in \sumtorus\ then reduces to the
computation in the above and we see that the
constraints $\omega_1 \simeq \omega_2$ arises upon the
summation. In general, when the ratios are
not necessarily rational, we can find
a series of rational numbers which converges to
$\theta_i/\varphi_i$. At each step in the series,
the sum over the coloring in \sumtorus\ gives the constraints
$\omega_1 \simeq \omega_2$. Thus it should also be the
case in the limit of the series.

This result is extended to surfaces of higher genera as follows.
A genus-$g$ surface $\Sigma$ can be constructed
from a $4g$-sided polygon by gluing its
sides together as is indicated in Fig. 14. Correspondingly
the surface is decomposed into $4g$ triangles.
As in the case of the torus,
we can parametrize the flat connection $\omega$ on $\Sigma$ by its
holonomies $U_{(a)}, V_{(a)}$
($a=1,...,g$) around the homology cycles $\alpha_a$ and $\beta_a$
as marked in Fig. 14.
These holonomies are subject to the constraint,
$$
    U_{(1)} V_{(1)} U_{(1)}^{-1} V_{(1)}^{-1}
    \cdots U_{(g)} V_{(g)} U_{(g)}^{-1} V_{(g)}^{-1}
        = {\bf 1}
\eqn\constraint
$$
In this case,
the wave-function $\Psi_{\Delta,c}(\omega)$
is a product of $U_{(a)}$'s and
$V_{(a)}$'s connected by the $3j$-symbols.
Especially it depends on $U_{(1)}$ as
$$
\eqalign{ \Psi_{\Delta,c}(\omega) = & \sum_{m_i,m_i',m_i''}
            U_{(1)j_1m_1'}^{~~~m_1} W^{m_3m_4} \cr
       &~~~~~~~~\times g_{j_1}^{m_1'm_1''} g_{j_2}^{m_2'm_2''}
          \cg j_1/j_2/j_3/m_1/m_2/m_3/
          \cg j_1/j_2/j_4/m_1''/m_2'/m_4/, \cr}
$$
where $W^{m_3m_4}$ is independent of
$U_{(1)}$. The sum of
$\Psi_{\Delta,c}(\omega_1)\Psi_{\Delta,c}(\omega_2)$
over $j_1$ and $j_2$ then imposes the constraint
$U_{(1)1}=U_{(1)2}$ as in the case of the torus.
The rest of the summation can be done inductively, and
we obtain the constraint $\omega_1 \simeq \omega_2$.

We have found that
the left hand side of \innertwo\ is equal to
$K(\omega_1,\omega_2)$ upto a constant factor $B_g$
$$
  \eqalign{&
   \sum_{c_1 \in C(\Delta_1)
           \atop c_2 \in C(\Delta_2)} \Psi_{\Delta_1, c_1}(\omega_1)
                    \Lambda^{-n(\Delta_1)} P_{\Delta_1,\Delta_2}(c_1,c_2)
          \Lambda^{-n(\Delta_2)}
            \Psi_{\Delta_2, c_2}(\omega_2) \cr
       &~~~~~~~~~~~~~~~~~~~~~~~~~~~~~~
       = B_g \cdot K(\omega_1,\omega_2) . \cr}
$$
Equivalently
$$
   ( \Psi_{\Delta_1,c_1}, \Psi_{\Delta_2,c_2})_{CS}
    = B_g^{-1} \cdot P_{\Delta_1,\Delta_2}(c_1,c_2).
$$
By combining this with \equivalence\ and by using the expressions
\cspartition\ and \prpartition , we obtain
$$
   Z_M^{(CS)} = A_g^2 B_g^{-1} Z_M.
\eqn\identity
$$
Now we would like to show that
$B_g$ is equal to $A_g^2$.
Since $A_g$ comes from the normalization
of the functional integral \hawking,
we must specify it in order to relate $B_g$ to $A_g$.
Here we define the normalization in such a way that
the Chern-Simons partition function for $S^3$ is equal to $1$.
To show $A_g^2B_g^{-1}=1$, we note that $S^3$
can be constructed from two handlebodies of any genus $g$.
Let us take a closed surface $\Sigma$ of genus $g$
and embed it into $S^3$. It is easy to see that
both the interior and the exterior of $\Sigma$ are
handlebodies of genus $g$. In this setting,
the left hand side of \identity\ is equal to $1$ due to
the normalization convention of $Z_M^{(CS)}$.
On the other hand, it follows from the definition of the lattice
gravity that $Z_M$ for $M=S^3$ is also $1$.
In this way, we have shown $A_g^2B_g^{-1}=1$ for any
value $g$. This proves the
equality of the partition functions of the lattice gravity
and the Chern-Simons theory.

Actually, there is a cavear here.
In the above, we have assumed that the integral \cspartition\
and the sum \prpartition\ are convergent. This is not always
the case. For example, when $M$ is of the topology $\Sigma \times S^1$,
the partition function $Z_M$ of the lattice gravity is given by
a trace over physical states
$$
    \sum_{c \in C(\Delta)} \Lambda^{-n(\Delta)}
                  P_{\Delta,\Delta}(c,c),
$$
where $\Delta$ is a triangulation of $\Sigma$.
Namely the partition function $Z_M$ counts the
number of physical  states for $\Sigma$ which is infinite if
$g \geq 1$. In this case,
the partition function $Z_M^{(CS)}$ for the Chern-Simons theory
also diverges as was pointed out by Witten [\Witten].
There he has shown that the divergence occurs when
$e \rightarrow \infty$, namely when the size of $M$ is large,
and thus it is infrared in nature.

So far, we have considered the case of a closed three-dimensional
manifold $M$. It is also possible to consider a manifold with
boundaries and discuss transition amplitudes between initial
and final states.

We have already studied some of such processes in this paper.
For example, $P_{\Delta_1,\Delta_2}(c_1,c_2)$ corresponds to
the geometry $N \simeq \Sigma \times [0,1]$ for
a transition of $\Sigma$ into $\Sigma$.
The Hartle-Hawking-type wave-function $Z_{\Delta,M}(c)$
can also be viewed as describing a transition of a point into
a closed surface $\Sigma$ (or a creation of $\Sigma$
from nothing). In both of these cases, we have found that
the amplitudes of the lattice gravity and the Chern-Simons
theory are related by the isomorphis defined in Section 2.

We can extend this analysis and study more elaborated
transition processes involving a topology-change of
$\Sigma$. In [\Witten], Witten has examined
the following situation in the case of the Chern-Simons
theory. Consider
$\Sigma_{initial}$
consisting of two components $\Sigma_1$ and $\Sigma_2$ of
respective genus $g_1$ and $g_2$. One can
construct a manifold $M$ which interpolates $\Sigma_{initial}$ to
another surface $\Sigma_{final}$
of genus $g=g_1+g_2$ consisting of
a single component, as follows. We first embed $\Sigma_{final}$
into ${\bf R}^3$ to obtain a handlebody $M_0$ of genus $g$.
We then remove from $M_0$
handlebodies of genus $g_1$ and $g_2$ whose boundaries are
$\Sigma_1$ and $\Sigma_2$. The remaining protion of $M_0$
gives the manifold $M$ whose boundaries are $\Sigma_1$, $\Sigma_2$
and $\Sigma_{final}$. A wave-function for the initial surface
$\Sigma_{initial}$ is spanned by products of functions $\Phi_{\Sigma_1}$
and $\Phi_{\Sigma_2}$ of flat connections on $\Sigma_1$ and
$\Sigma_2$. The transition amplitude for $M$
then relates $\Phi_{\Sigma_1}\Phi_{\Sigma_2}$
to $\Phi_{\Sigma_{final}}$ for the final surface. It is shown in [\Witten]
that the relation is as follows.
$$
   \Phi_{\Sigma_{final}}
                \sim \delta(A-{\bf 1}) \Phi_{\Sigma_1}\Phi_{\Sigma_2}.
\eqn\branching
$$
Here $A$ is an element of $SO(3)$ given in terms of holonomies
$U_{(a)}$ and $V_{(a)}$ ($a=1,..,g$) on $\Sigma_{final}$ as
$$
    A =  U_{(1)} V_{(1)} U_{(1)}^{-1} V_{(1)}^{-1} \cdots
             U_{(g_1)} V_{(g_1)} U_{(g_1)}^{-1} V_{(g_1)}^{-1},
$$
and the holonomies  $U_{(1)},...,U_{(g_1)}$ and
$V_{(1)},...,V_{(g_1)}$ correspond to the homology cycles on
$\Sigma_{final}$ which are homotopically equivalent
to the cycles on $\Sigma_1$ through the manifold $M$.

For the lattice gravity, the transition amplitude
$Z_{\Delta_1,\Delta_2;\Delta_{final}}(c_1,c_2;c_{final})$
for $M$
is given by summing over coloring of tetrahedra interior of
$M$ while keeping fixed the colorings $c_1$, $c_2$ and $c_{final}$
on $\Sigma_1$, $\Sigma_2$ and $\Sigma_{final}$.
To see its relation to the transition amplitude \branching\
in the Chern-Simons theory,
we multiply the Wilson-line
networks to $Z_{\Delta_1,\Delta_2;\Delta_{final}}$
and sum over the colorings as
$$
\eqalign{\sum_{c_1 \in C(\Delta_1), c_2 \in C(\Delta_2)
             \atop C_{final} \in C(\Delta_{final}) }
     & Z_{\Delta_1,\Delta_2;\Delta_{final}}(c_1,c_2;c_{final}) \cr
    & \times \Lambda^{-n(\Delta_1)} \Psi_{\Delta_1,c_1}
             \Lambda^{-n(\Delta_2)} \Psi_{\Delta_2,c_2}
             \Lambda^{-n(\Delta_{final})}
        \Psi_{\Delta_{fianl},c_{final}} . \cr}
$$
The computation is essentiall the same as we did in \innertwo ,
and the result agrees with \branching.
Thus the transition ampltudes of this type
are also equivalent in the two theories.

\section{Discussions}

We have found that the partition functions and some topology-changing
amplitudes of the lattice gravity of Ponzano and Reggeare equal
to the ones of the $ISO(3)$ Chern-Simons theory. This result supports
the original conjecture by Ponzano and Regge that the statistical sum
of the $6j$-symbols describes the fluctuating geometry with the
weight $\prod_{x \in M} \cos(\sqrt{g} R)$. Indeed, if we integrate over
$\omega$ first in the Chern-Simons functional integral,
we are left with an integral over the dreibein $e$ with a weight
$\exp (\int e \wedge R)$ where $R$ is a curvature two-form constructed
from $e$.
Although we seem to have gotten $\exp $ rather than $\cos $,
we should note that $\int e \wedge R$ changes its sign if we flip
the orientation of $e$. Since we integrate over $e$ as a part of
the $ISO(3)$ gauge field, at each point in $M$,
both orientations of $e$ contribute to the functional integral. If
one tries to integrate over $e$ of a fixed orientation,
one would need to replace the exponential by the cosine to
compensate for the restriction.
Therefore it appears that the lattice gravity
of Ponzano and Regge gives a functional integral of
$\prod_{x \in M} \cos( \sqrt{g} R)$ with the correct measure
for the fluctuating metric on $M$.

To regard this system as the Euclidean Einstein gravity,
the factor $i$ in front of the action is disturbing.
One cannot eliminate it by rotating the contour of the
$e$-integral since the resulting functional integral would
be divergent. To address this issue, it would be more fruitful
to study a Lorentzian version of the lattice model based on
the infinite dimensional representations of $SO(2,1)$.
Since the representation theory of $SO(2,1)$ is far richer
than that of $SO(3)$, we must have good criteria in
choosing a class of representations we
put of the edges of the tetrahedron.
One of the criteria would be that a sum of characters
over such representations should give the invariant $\delta$-function
$\delta(U - {\bf 1})$, in order for the lattice model to have
the same partition function as in the Lorentzian Chern-Simons gravity.
Such study will be useful in understanding the structure of
physical observables in the Lorentzian gravity.

Recently Mizoguchi and Tada \REF\MT{S.~Mizoguchi and T.~Tada,
Kyoto preprint,
YITP/U-91-43 (1991).}[\MT] have studied the
$q$-analogue of the $6j$-symbol and found the
asymptotic formula for $q = e^{2 \pi i/(k+2)}$
$$
   (-1)^{\sum_i j_i} \wsymbol j_1/j_2/j_3/j_4/j_5/j_6/_q
    \sim {c \over \sqrt{V}}
      \cos (S_{Regge} - {\lambda_k \over 2} V + \pi./4),
$$
where $c$ is some constant and $\lambda_k = (4 \pi/k)^2$.
Thus it is natural to expect that the $q$-analogue of the
Ponzano and Regge model introduced by Turaev and Viro
would be related to the gravity with
the cosmological constant $\lambda_k$ or the
$SO(3) \times SO(3)$ Chern-Simons theory. The recent paper
by Turaev [\Turaev] supports the latter possibility.

Durhuus, Jakobsen and Ryszard \REF\Dur{B.~Durhuus, H.P.~Jakobsen
and R.~Nest, ``{\it Topological Quantum Field Theories
from Generalized 6j-Symbols,}'', preprint (1991).}[\Dur]
have constructed a large class of topological lattice models
extending the model of Turaev and Viro. The method developed in
this paper could also be applicable to study those models.

\vskip 3mm
\noindent
{\bf Acknowledgments}

The author would like to thank N.~Sakakura for discussions.
He is also thankful to T.~Maskawa for explaining him
some group theoretical facts and to L.~Kauffman for encouragements.

\refout

\endpage

\centerline{\bf FIGURE CAPTIONS}

\noindent
1. A tetrahedron colored by angular momenta $j_i$'s.

\noindent
2. One tetrahedron can be decomposed into four small tetradedra.

\noindent
3. A manifold $M$ is decomposed into three parts,
$M_1$, $M_2$ and $N$.

\noindent
4. $Z_{M_i,\Delta_i}(c_i)$ is defined by the summation
over colorings on edges interior of $M_i$ with the fixed coloring $c_i$
on the boundary.

\noindent
5. Attaching a tetrahedron to a triangle on a surface, as seen from the
above, and the corresponding Wilson-line networks.

\noindent
6. Attaching a tetrahedron to two neighbouring triangles on a
surface, as seen from the above, and the corresponding Wilson-line
networks.

\noindent
7. A genus-$g$ surface can be constructed from two discs with
$g$-holes by gluing their boundaries together. The case of
$g=3$ is shown in the figure.

\noindent
8. Part of Wilson-lines on $S_g^+$ and $S_g^-$.

\noindent
9. The Wilson-line $j_3$ in Fig. 8 can be removed using the
orthogonality of the $3j$-symbols.

\noindent
10. Triangulation of $S_{g=1}^\pm$ in the case of a torus.

\noindent
11a. Wilson-line network corresponding to the triangulation in Fig. 10.

\noindent
11b. The Wilson-line $j_2$ is removed using the orthogonality of the
$3j$-symbols.

\noindent
11c. The Wilson-line $j_1$ can be moved around using the flatness of
$\omega_{|\Sigma}$.

\noindent
11d. The homotopically trivial loop is removed.

\noindent
11e. The network is transformed into a single Wilson-loop around
the homology cycle contractable in $M$. The sum over $j_4$
restricts the holonomy around this cycle to be trivial.

\noindent
12a. Triangulation of $S_g^\pm$.

\noindent
12b. The corresponding Wilson-line network on $S_g^\pm$.

\noindent
13a. Torus can be decomposed into two triangles.

\noindent
13b. The corresponding Wilson-line network.

\noindent
14. A genus-$g$ surface can be constructed from a $4g$-sided polygon.
Correspondingly, the surface is decomposed into $4g$ triangles.

\bye
\end